# A Voice-based Triage for Type 2 Diabetes using a Conversational Virtual Assistant in the Home Environment

Kelvin Summoogum
*CEO & Founder*
*MiiHealth*
Arizona, USA
kelvin@miihealth.ai

Debayan Das
*Data Scientist*
*MiiCare Ltd*
London, UK
debayan.das@miicare.co.uk

Sathish Kumaran
*Clinical Data Analyst*
*MiiCare Ltd*
Bangalore, India
sathish.kumaran@miicare.co.uk

Dr. Sumit Bhagra, MD
*Chair of Endocrinology*
*Mayo Clinic Health System*
Minnesota, USA
bhagra.sumit@mayo.edu

*Abstract* — Incorporating cloud technology with Internet of Medical Things for ubiquitous healthcare has seen many successful applications in the last decade with the advent of machine learning and deep learning techniques. One of these applications, namely voice-based pathology, has yet to receive notable attention from academia and industry. Applying voice analysis to early detection of chronic diseases holds much promise to improve health outcomes and quality of life of patients. In this paper, we propose a novel application of acoustic machine learning based triaging into commoditised conversational virtual assistant systems to pre-screen for onset of diabetes. Specifically, we evaluate a novel triaging system which extracts acoustic features from the voices of n=24 older adults when they converse with a virtual assistant and predict the incidence of Diabetes Mellitus (Type 2) or not. Our triaging system achieved hit-rates of 70% and 60% for male and female older adult subjects, respectively. Our proposed triaging uses 7 non-identifiable voice-based features and can operate within resource-constrained embedded systems running voice-based virtual assistants. This application demonstrates the feasibility of applying voice-based pathology analysis to improve health outcomes of older adults within the home environment by early detection of chronic conditions like diabetes.

## I. Introduction

Several medical conditions have been shown to affect the voice due to changes that either occur in the speaker's speech organs or in the brain. These changes in voice could theoretically be analysed to learn specific acoustic patterns, and generalised across populations, irrespective of languages spoken. There have been prior publications in the clinical literature over the past decade examining voice patterns in different neurological and psychological conditions like Stress [1], Autism [2], Dysphonia [3], Parkinson's Disease [4], Alzheimer's Disease [5], COVID19 [6], etc. Of note, a "digital biomarker" is a characteristic objectively measured by a digital device or captured by a digital algorithm that is evaluated as an indicator of normal biological processes, pathogenic processes or pharmacologic response to therapeutic treatments. A voice-based digital biomarker in this regard, converts human speech audio input into an evaluation of subject's speech generation capabilities. Voice-based (vocal) digital biomarkers are alluring and well-suited features for predictive algorithms like that in machine learning and deep learning as they are non-invasive, instantaneous and cost-effective to use and easy to scale across large populations. There is a pressing need for innovative solutions in medicine to better understand diseases, improve health outcomes, and alleviate the growing constraints of health care budgets. However, the application of vocal biomarkers has been largely limited to the neurological diseases [7], per our review of clinical literature. In contrast, the use of vocal biomarkers remains scarce.

We noted that almost all of the existing works using voice biomarkers to predict for pathologies use audio data collected in controlled acoustic scenes [3, 8, 9, 10, 11, 12, 13] with little to no ambient noise and presence of overlapping acoustic events. The authors of [12] collected speech data by having the participants say pre-written words to reduce influence of language during the analysis. The data-capture primarily happened several times within a well-defined data collection period. We were unable to find relevant literature where voice biomarkers were collected continuously to showcase changes over periods longer than a few weeks [10, 13, 14] or days [8, 12, 13]. Such methodologies for voice biomarkers are restrictive by design and do not lend into ease of integration within existing care pathways in real-world deployments. Similarly, accessibility by vulnerable demographics like older adults and those with high social deprivation [15], disadvantageous social determinants of health [16] (SDOH) and individuals with cognitive challenges, further influence their feasibility. It is precisely this challenge that drives our motivation to develop a new approach to collect voice biomarkers in a more sustainable manner. Our goal is to enable data-capture through widely accepted Assistive Technologies (AT) that already rely on voice-based conversations for engagement and care delivery, ensuring seamless integration into existing systems of care.

Conversation-based AT fall within the domain of "telehealth" which promises the advantage of enabling patients to self-care by delivering supportive care, guidance and necessary health literacy, remotely [17, 18, 19]. The COVID19 pandemic gave strong impetus [17] to the rise of telehealth tools and saw rapid adoption of programmable conversational agents [18] and virtual assistants [19] like chatbots and voice-based assistants. Since 2020, there has been several published literature reviews demonstrating [17, 18, 19] that virtual assistants and chatbots can be used to assist in clinical workflows, deliver health programs, personalise support at scale with low costs.

In this paper, we investigate the feasibility of using a commercially available conversational virtual assistant product to capture speech audio from community-dwelling older adults, with the aim of extracting voice biomarkers that strongly correlate with a diagnosis of type 2 diabetes mellitus and predicting the same for the individual. We also envision our approach as a "triaging toolkit" that can be adapted to predict other conditions [1, 2, 3, 4] affecting the voice, such as high stress, mental disorders, neurodegenerative conditions and non-communicable diseases [8, 10, 12]. However, these applications are beyond the scope of this study. We summarise the contributions of this work as follows:

1. Using virtual assistants to perform voice-based triaging for type 2 diabetes, one of the most prevalent [20] medical conditions across the world.
2. Developing a "triaging toolkit" suited for embedded system applications to comply with data-protection and user-privacy regulations like GDPR and HIPPA.
3. Improving triaging robustness in voice biomarker extraction in noisy polyphonic environment of the home environment.

We have organised the paper as follows: In Section II, we present the current state-of-the-art of voice biomarkers, its extraction from voice-based sources, diagnostic relationships drawn between these biomarkers and type 2 diabetes, and underpinnings of virtual assistants in personalised care coaching and delivery. In Section III, we explain our approach leveraging Monica, a voice-based conversational health assistant developed by MiiCare [21]. In Section IV, we lay out the results of our method applied to a cohort of older adults where Monica had already been deployed to support daily chronic management.

## II. BACKGROUND

### A. Voice Disorders

Voice and speech production involves the coordinated functioning of multiple organ systems. Airflow generated by release of pressure from the lungs passes through the vocal folds in the larynx, causing them to vibrate which produces sound. The articulation of this sound forms speech. [22]. This functional dependency between several biological structures makes the voice vulnerable to being affected by diverse conditions.

Anomalies in vocal quality, such as pitch, volume,, resonance and duration, that are unexpected for individuals regardless of their gender and age, are indicative of voice disorders. There is no globally accepted nomenclature for voice disorders. The authors of [22] list structural, inflammatory, traumatic, systemic, aerodigestive, psychiatric and psychological, neurological and functional voice disorders as major categories in this subfield. Medical specialists (like a Speech Therapist) can diagnose voice disorders through several in-person examinations and tests. The current approach relies on clinical examinations consisting of interviews, perceptual voice evaluation, patient-reported outcomes, laryngoscopy, aerodynamic assessment, voice profiles, acoustic spectral analysis, and laryngeal electromyography [22], which is time consuming, resource expensive and generates high economic burden. Appraising the vocal biomarkers gathered by a clinician is what ultimately leads to a formal diagnosis of a disorder. Pre-screening at source, for instance, within the home environment of the patient, not only reduces the overall cost of diagnosis, but also the workload on the healthcare system by prioritising the undiagnosed who are at higher risk.

### B. Blood Glucose and Voice

Diabetes mellitus is a chronic metabolic disorder characterized by the impaired regulation of blood glucose. It is broadly categorized into two types: type 1 diabetes, which results from inadequate production of insulin by pancreas, and type 2 diabetes, which is caused by reduced cellular sensitivity to insulin. Type 2 diabetes accounts for the vast majority of diabetes cases, comprising approximately 90-95% of all diagnosed cases, while type 1 diabetes represents a smaller proportion.

Authors of studies like [23, 24, 25, 26] make the following claims as to how voice is affected by blood glucose:

1. Glucose related changes in voice can be perceived. For example, the authors of [23] and [24] found that people who know a diabetic subject well can hear when they are hypoglycaemic, i.e. the blood glucose level drops below the normal physiological range.
2. Fluctuations in blood glucose levels can alter the elastic properties of the biological tissues in the larynx and the vocal cords, leading to changes in the spectral properties of the voice in individuals with diabetes. The authors of [25] and [26] put forward this idea in compliance with Hooke's law of physics.
3. Hypoglycaemia is often accompanied [27] by feelings of anxiety, causing people to speak faster and with greater urgency, whereas hyperglycaemia (higher levels of blood glucose over normal baseline) is often accompanied by fatigue or lethargy causing speech patterns to be slow or slurred.

Although no supporting references was found in these publications, our review found that these association between glucose levels and change s in speech patterns is plausible and supported by physiological principles. Subjective claims like detecting abnormal glucose levels from changes in voice has been reported since 1990s [28] and the similarity of symptoms in anxiety and hypoglycaemia [29]. Literature using Hooke's law or changes in elasticity of the larynx and vocal cords to explain changes in voice is sparse. The earliest work detected by the authors of [25] on this argument provided experimental evidence that increase in blood glucose levels reduce elasticity in the muscles of the aforementioned organs. Authors of [8, 10, 11, 12] recently investigated this hypothesis and found that glucose levels and the fundamental frequency (F0) [30] of the voice had significant positive correlation per subject. They also reported that this effect was observed across subjects even when grouped by their diagnosis of non-diabetic, pre-diabetic, and type 2 diabetes).

### C. Current State of Voice-based Virtual Assistants

Intelligent conversational agents and virtual assistants proved [17] their potential to benefit over-burdened healthcare systems and provide personalised care to patients within the comfort of their own homes during the COVID19 pandemic through their versatility, accessibility, scalability for naturalistic communications with them. Arguably, the popularity of consumer facing virtual assistants may have started with the launch of Apple Siri in 2011. After Amazon introduced Alexa and Echo devices in 2014, the technology around automatic speech recognition, text-to-speech and speech-to-text, natural language processing methods, and conversational AI has been continuously improving as the number of users has been increasing [17, 18]. Recently, with the introduction of GPT models [31], Large Language Models (LLM) [32] have taken over this field as the state-of-art for any conversational virtual assistant technology due to its ability to conduct extremely realistic human-like conversations in real time. Virtual assistants have been adopted [17] by healthcare institutions for applications such as healthcare literacy, to respond to health information, tips and guidelines requested by users, health news, updates about hospital operations, first aid guidance, and medical communications. The convenience of natural conversation and hands-free interaction through digital devices has the potential to improve the effectiveness of health information delivery and communication.

*D. Current State of Predictive Analysis of Diabetes*

The traditional diagnostic methods for diabetes, while effective, often diagnose diabetes after significant metabolic disruption has occurred. Therefore, there is a growing emphasis on developing earlier dictation technologies to prevent complications an implement timely therapeutic intervention. Advances in machine learning provide several promising avenues for early detection. The authors of [33] proposed stacking-based ensemble classification methods, which integrate outputs from various algorithms to improve predictive accuracy, to predict type 2 diabetes. This approach has been applied using the publicly available PIMA Indian diabetes dataset to predict the onset of Type 2 diabetes within a 5-year interval [34]. The authors of [35] presented a soft computing-based diabetes prediction that uses the PIMA Indian dataset alongside breast cancer datasets for evaluation purposes. Both publications rely on traditional machine learning algorithms like SVM, Decision Tree, RBF SVM, Logistic Regression and Naïve Bayes. The authors of [35] were able to achieve a cross-validated accuracy of 79%. The authors of [36] used the learnings from the work of [35] and applied the algorithms on a dataset of 529 individuals from a hospital in Bangladesh through questionnaires. Their experimental results show that Random forests tend to outperform compared to other algorithms. A review of machine learning and statistical learning methods in publications produced between 2010 – 2020 [37] indicates that the Random Forest algorithm consistently outperforms other machine learning mild, including those based on neural network, particularly when applied to various tabular datasets. All of these works use demographic and patient health data to predict the subject's risk of diabetes.

Recently, there has been renewed interest in predicting whether a subject's blood glucose levels are high or low by analysing their voice audio. The authors of [8] positions speech audio as a "biomarker sensor" by applying machine learning classification. They achieved a $f(1)$ score of 0.88 in a leave-one-out cross-validated dataset of 70 audio samples per subject with 49 participants in total. The authors of [10] investigated whether voice analysis can be used as a pre-screening tool for type 2 diabetes by examining the difference in voice recordings between individuals with diabetes and individuals without diabetes. They found significant differences, both in the entire dataset and in age-matched and BMI-matched samples and their predictive models achieved $f(1)$ scores of 0.75 for women and 0.7 for men separately. Both works [8, 10] found that frequency-based voice features like pitch, jitter, shimmer, etc. show high predictive power and collected data samples via a smartphone-based voice recorder.

### III. METHODOLOGY

Our method focuses on collecting voice biomarkers from regular everyday engagements between a voice based virtual assistant and patients who are primarily community-dwelling older adults, living alone. In this section, we present the end-to-end pipeline we developed for data collection and application of a machine-learning based classifier in a triaging system for the continuous screening of the onset of type 2 diabetes.

*A. Research Motivations*

The works of [8, 10, 12, 33-37] have shown that it is feasible to use machine learning algorithms to analyse voice features and classify them to belong to someone with diabetes or not. Applications of such models in real world clinical practice has however been challenged [38] regarding reliability and utility of its predictions. The existing works focus on individual sample classification performance but fails to demonstrate how such models can be used for early intervention where its utility is maximised. These works rely exclusively on discrete collection of feature samples and is not clear how they can integrate within the daily routine of older adults who may not be comfortable using smartphones or similar data-collection devices. Our motivation behind this paper was to develop a triaging system instead of a simple classifier, which can be:

1. seamlessly integrated with voice-based virtual assistants which do not require digital literacy to use.
2. used alongside assistive technology in clinical and social care pathways like virtual wards [39], hospital-at-home [40], and discharge-to-assess [41]
3. used to prioritise those who are at high risk to avoid adding unnecessary burden on healthcare systems
4. flag at first signs of onset, to intervene early improve patient outcomes.

*B. System Design*

To support our motivation, we put forward the following conceptual design of our triaging system in Figure 1 below.

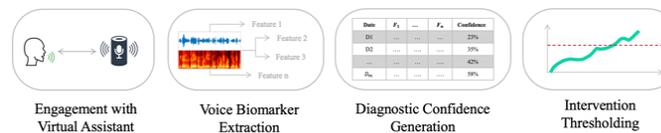

*Figure 1 Conceptual Flowchart of Our Proposed Triaging System*

The system has 4 steps in implementation:

1. **Voice Capture:** A home hub which hosts the virtual assistant is deployed in the home of an older adult patient who needs support and long-term care monitoring. The assistant interacts with the patient every day and collects their responses as voice recordings.
2. **Biomarker Extraction**: The home hub runs a program that analyse the voice recordings and extract relevant acoustic and vocal features required to ascertain risk. The program only sends non-identifiable features to the cloud in accordance with user privacy regulations like GDPR and HIPPA.
3. **Diagnostic Claim Generation:** A machine learning model is trained and accepts specific voice features to output the probability that these features are similar to those on the trained dataset of features from people at risk. The model generates probability values accordingly, for all feature sets captured every day and averages them to give a confidence percentage or score for the older adult is at risk.
4. **Intervention Thresholding**: The daily confidence scores are then monitored over a period of time. Medical expertise will advise how long the data will need to monitored and what confidence threshold criteria will need to be met before the patient will be deemed to be at risk and require a formal intervention from the healthcare system.

In this paper, we focus only on the first 3 steps, and we discuss the method we followed to train a machine learning model and use it to generate confidence scores for tracking onset of diabetes. The last step pertaining to threshold confidence scores for intervention is out of scope of this paper and will be covered in a future publication.

*C. Sourcing a Voice-based Virtual Assistant*

The first step in our triaging system required a virtual assistant technology. We needed a voice-based conversational assistant which would not only help us to collect speech voice samples but also provide value to the older adult subjects. The works of [42, 43] have noted that older adults are more likely to adopt new digital technology as part of their daily life if it actively supplements their regular needs.

We sourced Monica (abbreviated for "MiiCare's On-demand Neural Intelligence Care Assistant"), a new voice-based conversational virtual assistant from MiiCare [21], marketed as a virtual companion and health coach for older adults who live alone and have chronic management needs. Monica is housed within the MiiCube, a small cube-shaped device that serves as a home hub controlling a wireless remote patient monitoring ecosystem within the home environment.

The MiiCube uses a ARM4 processer coupled with a 2-microphone setup from Seed Studio's ReSpeaker Pi Hat [44] enclosed in a cubical box enclosure as shown in Figure 2. The microphones are located immediately beneath the top grill to ensure it is capturing the acoustic scene and all acoustic events including the patients' speech responses without exposing the computing unit.

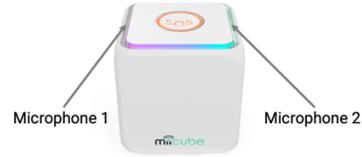

*Figure 1 The MiiCube recording device from MiiCare*

To comply with user privacy regulations, the device does not background record audio. It captures the audio recording of the patient's responses only when either Monica or the patient initiates a conversation. In this study, we record the patient's response at each turn of the conversation and for a maximum of 10s, to minimise unnecessary recording. The speech audio is sampled at 16 KHz 16-bits in PCM WAV format and immediately processed within the MiiCube to return a response. Monica uses Azure OpenAI [45]'s GPT inference to provide human-like natural conversational experience for the older adult subjects as well as to provide personalised content in the form of reminders and behavioural prompts to support self-care and condition management.

*D. Voice Biomarker Extraction*

As part of this research study, we modified the MiiCube to run a Python-based script alongside its main system program. This Python script accepts a copy of the speech recording data captured by the MiiCube system program and extracts voice biomarkers. The reasoning behind running the voice biomarker extraction step at the point of collection instead of in the cloud is to comply with ethics review and ensure highest levels of user privacy and data leak protection. This implementation demonstrates regulatory compliance by design, for privacy-sensitive applications like in clinical trials and even the commercial mass market.

The MiiCube's compute environment runs Python 3.7.3, which can only support 32-bit operating systems. The authors of [8, 10, 11, 12] had used Python modules like Librosa [46], praat-parselmouth [47] and disvoice [48] to process voice audio files and extract voice biomarkers. Librosa cannot be used in 32-bit compute environments, therefore we used scipy.utils [49] instead to load the voice audio recordings and praat-parselmouth to extract the acoustic features as they were supported by our compute environment.

We extracted 7 acoustic features from the voice recordings using praat-parselmouth. Some of these acoustic features have already been demonstrated to have strong correlation with blood glucose levels in the works of [8, 10, 11, 12, 23, 24, 25, 26]. We list these features in Table 1 below.

TABLE I. Acoustic features used as voice biomarkers for type 2 diabetes

| Feature name | Description | Citations |
|---|---|---|
| Articulation Rate | Syllables spoken per second, excluding pauses. The average rate for English speakers is 6.19 syllables per second | [47, 50] |
| Speaking Rate | Words spoken by a speaker within a minute to deliver information. Slow casual conversations are usually between 120 – 150 words per minute. | [47, 50] |

| | Frequency variation in voice pitch, measuring cycle-to-cycle variations in the fundamental frequency during phonation. Expressed as a percentage. | |
|---|---|---|
| Jitter | Frequency variation in voice pitch, measuring cycle-to-cycle variations in the fundamental frequency during phonation. Expressed as a percentage. | [30, 47, 50, 51] |
| Shimmer | Amplitude variation in the loudness of voice. Measures cycle-to-cycle variations in amplitude during phonation. Expressed either as a percentage or in decibels (dB). | [47, 50, 52] |
| $\mu(F0)$ | Mean approximation of the fundamental frequency of voice | [47, 50] |
| $\sigma(F0)$ | Standard deviation of the approximation of fundamental frequency of voice | [47, 50] |
| $\sigma^2(F1)$ | Variance in the first formant frequency associated with vowel quality and resonance of the vocal tract during speech production. This frequency contributes to the perceived sound quality of vowels, specifically their open-ness and closedness during phonation. | [47, 50, 52] |

*E. Data Collection*

To collect the acoustic features from the MiiCubes, we built an API endpoint connected with a noSQL database on Azure Cloud. The API endpoint was written using Python and FastAPI [53] to accept 1x7 floating-point row vectors along with the subject identifier and MiiCube device identifier. This API accepted each feature vector and stored them in an Azure Cosmos DB in an unstructured JSON Collection. This step is unique to our setup, as far as we are aware, and can be replaced with any suitable data collection and storage suitable with voice assistant devices. Storing the data using this way made it easy for us to add or remove subjects to the research in a staggered fashion without worrying about the data collection.

MiiCare received consent from n=50 older adult MiiCube users living in the UK to be part of this study. Our python script, written to extract the 7 acoustic features, was remotely installed in these MiiCubes. We let the script run continuously for 30 days across these 50 MiiCubes for the entire duration of October 2024. All these users shared the following similarities:

1. The subject lives alone and can self-care.
2. The subject resides in the UK and speaks English as their main language.
3. The subject takes one or more medications as part of their daily chronic management needs.
4. The subject had consented to have Monica be pro-active in its engagement approach to collect more voice recordings from them.

We set the system to collect 6 voice recordings over 3 separate days, i.e. 2 recordings per day at least. We have noted that it is possible for a subject to have several conversations with Monica and thereby collect more than 2 recordings in a single day. The acoustic features in these recordings is influenced by the varying blood glucose levels throughout the day due to factors like meal timings, physical activity, stress levels, etc. and this holds for diabetic as well as non-diabetic subjects. We do not want our triage to be impacted by this and by setting a minimum data collection duration of 3 days, we minimise the influence of temporary blood glucose fluctuations on the triage's performance to track onset of diabetes. There was no attempt undertaken to introduce any strict daily routine because we wanted to demonstrate feasibility of the method in real-world applications where the subjects are not expected to change their daily routine to accommodate the assistive technology.

Monica was setup to support the older adult users with their daily chronic management needs. This included daily reminders for timely medication and fluid intake. In addition to this, Monica was programmed to initiate engagement with the user at home when they are idle. Users were free to chat with Monica whenever they wanted. The setup is similar to real world deployments of smart home assistants and virtual agents like Alexa, Siri and Google Home when used in healthcare applications [17, 54].

*F. Dataset*

After 30 days, we took stock of all the features collected for each subject from our cloud database and applied our sampling threshold of 2 recordings every day. We found that n=24/50 MiiCube users had enough data to satisfy this threshold. The reasons for the remaining 26 MiiCubes failing to collect enough data was attributed to poor internet connectivity as those patients lived in rural areas. We summarise the feature samples collected for each user in Table II below. In total, we had data from n=17 male users (11 with diabetes and 6 without diabetes) and n=7 female users (3 with diabetes and 4 without diabetes).

TABLE II. Acoustic feature samples collected for n=24 MiiCube users

| Gender | User ID | Feature samples | Days |
|---|---|---|---|
| male | n=11 users with confirmed diagnosis of type 2 diabetes | | |
| | ID-M1 | 396 | 28 |
| | ID-M2 | 201 | 30 |
| | ID-M3 | 101 | *28* |
| | ID-M4 | 89 | 20 |
| | ID-M5 | 83 | 24 |
| | ID-M6 | 59 | 20 |
| | ID-M7 | 29 | 15 |
| | ID-M8 | 22 | 11 |
| | ID-M9 | 16 | 8 |
| | ID-M10 | 12 | 6 |
| | ID-M11 | 11 | 7 |
| | n=6 users without diabetes | | |
| | ID-M12 | 181 | 25 |
| | ID-M13 | 163 | 28 |
| | ID-M14 | 147 | 29 |
| | ID-M15 | 73 | 15 |
| | ID-M16 | 42 | 11 |
| | ID-M17 | 30 | 13 |
| female | n=3 users with confirmed diagnosis of type 2 diabetes | | |
| | ID-F1 | 56 | 17 |
| | ID-F2 | 29 | 16 |
| | ID-F3 | 17 | 4 |
| | n=4 users without diabetes | | |
| | ID-F4 | 194 | 18 |
| | ID-F5 | 113 | 9 |
| | ID-F6 | 62 | 19 |
| | ID-F7 | 40 | 10 |

*G. Preprocessing Voice Biomarkers*

Since the voice features captured were all continuous floating-point numbers, preprocessing them requires careful feature scaling which should have clinical underpinnings. We enlisted with the following considerations as part of the scaling process:

1. The authors of [50] defined normative ranges for articulation rate for human speakers across a variety of languages. English speakers have an articulation rate of 6.19 syllables per second on average across the globe. We defined the scaling as a percentage ratio of how fast the observed rate is relative to this average. In formulaic terms, if the articulation rate is $ar$, then the scaling rule is $ar_{scaled} = \frac{ar - 6.19}{6.19}$
2. The authors of [50] establishes the normative range of speaking in English speakers at different paces for different settings. We used the rate of 150 words per minute (WPM) as the upper bound for casual everyday conversation and applied the scaling rule for speaking rate $sr_{scaled}$ as $sr/150$.
3. The authors of [52] gave us the normative range for jitter in both males and females to be between 0.00106% and 0.02312%. We applied min-max scaling rule for jitter as follows $jitter_{scaled} = \frac{jitter - 0.00106}{0.02312 - 0.00106}$.
4. As per praat-parselmouth's documentation, shimmer is measured as a percentage and is already scaled. There is however no normative range established for voice shimmer for pathologic conditions. For normal human voices, the expected value is less than 5% [52]. We used this instead and applied max-scaling rule for shimmer as follows: $shimmer_{scaled} = \frac{shimmer}{0.05}$
5. As per praat-parselmouth's documentation, the fundamental frequency $F0$ is set between 75 Hz and 300 Hz for human beings. We again applied min-max scaling rule for $\mu(F0)$ and $\sigma(F0)$ as follows:
    a. $\mu(F0)_{scaled} = \frac{\mu(F0) - 75}{300 - 75}$
    b. $\sigma(F0)_{scaled} = \frac{\sigma(F0)}{300 - 75}$
6. As per praat-parselmouth's documentation, the first formant frequency $F1$ is set between 200 Hz to 1000 Hz for human beings. We therefore applied min-max scaling rule as $\sigma^2(F1)_{scaled} = \frac{\sigma^2(F1) - 200}{1000 - 200}$

*H. Training the Machine Learning model*

Taking inspiration from the work of [10], we adopted the leave-one-out (LOO) approach to train and cross-validate the performance of the machine learning model for ex-situ samples for a subject whose features the model hasn't been trained on. The LOO approach was applied separately for the male and female subjects.

The works of [8, 10, 11, 12, 23, 24, 25, 26] has used various shallow machine learning algorithms like SVM [55], Logistic Regression [56], KNN [56], Decision Trees [56] and Random Forests [56]. The authors of [37] have even claimed the shallow algorithms are superior to neural network based deep learning approaches as they demonstrate similar predictive power at lower computational complexity and resource needs. The authors of [10] concur with this observation and added that neural networks overfit on such datasets as these are sparse in nature with only a few hundred samples that could be collected per patient in real life deployments. We agreed with this analysis and therefore we only focused on evaluating the set of algorithms and choosing the best performing model using our acoustic features.

We describe the exact steps followed for evaluation as follows:
1. Choose 1 subject identifier $ID_{holdout}$ from among the subjects available.
2. Separate the feature samples for $ID_{holdout}$ from the other subject identifiers $ID_{others}$.
3. Train the machine learning model $M$ on the feature dataset for $ID_{others}$. Please note that the label for each feature set will be 1 if the subject has diabetes and 0 if they do not.
4. Once the training is complete, apply $M$ on the feature set for $ID_{holdout}$ and extract the probability values $\{p_1\}$ for label=1 only. We want to focus on how likely the model thinks each feature set from $ID_{holdout}$ has similar characteristics to that of features of subjects in $ID_{others}$ who had label=1 (with diabetes).
5. $\mu(\{p1\})$ gives us an averaged probability score that $M$ thinks the features of $ID_{holdout}$ is like those with diabetes. Please note that the feature set must be collected over a period of several days when this step is executed in a real-life inference application (3 days as a minimum as explained already in Subsection E). This will give assurance that factors affecting blood glucose levels is controlled for in the generation of the average probability score for diabetes within that period of time.
6. We now apply a decision framework to determine whether we make the claim $C$ that subject $ID_{holdout}$ has diabetic symptoms using $\mu(\{p1\})$. We reason that (a) if $\mu(\{p1\}) > 0.6$, then it is highly probable that the subject is diabetic and should be flagged for intervention ($C = 1$); (b) If $\mu(\{p1\}) < 0.4$, then the subject is less probable to be diabetic and therefore currently at low risk for the same ($C = 0$); (c) If $0.4 <= \mu(\{p1\}) <= 0.6$, then we are not confident enough at this stage to make a diagnostic claim about the subject ($C = -1$).
7. We repeat Steps 1 – 6 for all subjects available for a either gender separately and tabulate the following information: $ID_{holdout}$, $\mu(\{p1\})$, claim made using our decision framework $C$, and lastly whether $C$ matches the ground truth diagnostic label for $ID_{holdout}$ (1 if has diabetes and 0 otherwise).

Tracking the number of claims for $C = 1$ over time will show the probability of onset of diabetes in undiagnosed subjects when the triage is applied to real-life. Determining the threshold of positive claims needed and the number of days to track onset must be made in consultation with clinical experts and is outside of the scope of this paper, as mentioned earlier.

## IV. RESULTS

We choose to evaluate the following algorithms in our machine learning based triaging system. We include the specific parametric configurations we set for each algorithm [57] as well:

1. `sklearn.linear_model.LogisticRegression(max_iter =500,class_weight="balanced")`
2. `sklearn.tree.DecisionTreeClassifier(class_weight = "balanced")`
3. `sklearn.ensemble.RandomForestClassifier(n_estima tors = 500, class_weight = "balanced")`
4. `xgboost.XGBClassifier(n_estimators = 300)`
5. `sklearn.neighbours.KNeighboursClassifier(n_neigh bours = 5, weights = "distance")`
6. `sklearn.naive_bayes.GaussianNB()`
7. `sklearn.svm.SVC(kernel = "rbf", class_weight = "balanced", probability=true)`
8. `sklearn.svmSVC(kernel = "linear", class_weight = "balanced", probability=true)`

We compiled the information points extracted in each LOO run for each of the above algorithms for either gender as Tables III and IV below. The columns are defined as follows in terms of the number of claims $C$ made for each subject using our decision framework:

1. $C_{correct}$ : how many subjects were correctly diagnosed
2. $C_{incorrect}$: how many subjects were wrongly diagnosed
3. $C_{undecided}$: how many subjects could not be diagnosed
4. $Hit\ rate$: defined as $\frac{C_{correct}}{C_{correct}+C_{incorrect}}$

TABLE III. Diagnostic Performance using LOO on n=17 male subjects

| Algorithm | $C_{correct}$ | $C_{incorrect}$ | $C_{undecided}$ | Hit rate |
|---|---|---|---|---|
| RF | 7 | 3 | 7 | 0.7 |
| DT | 6 | 3 | 8 | 0.67 |
| KNN | 7 | 4 | 6 | 0.64 |
| XGB | 7 | 4 | 6 | 0.64 |
| SVM-RBF | 5 | 5 | 7 | 0.5 |
| GNB | 4 | 4 | 9 | 0.5 |
| SVM-Linear | 4 | 6 | 7 | 0.4 |
| LR | 0 | 2 | 15 | 0 |

TABLE IV. Diagnostic Performance using LOO on n=7 female subjects

| Algorithm | $C_{correct}$ | $C_{incorrect}$ | $C_{undecided}$ | Hit rate |
|---|---|---|---|---|
| DT | 3 | 2 | 2 | 0.6 |
| KNN | 4 | 3 | 0 | 0.571 |
| RF | 4 | 3 | 0 | 0.571 |
| SVM-Linear | 4 | 3 | 0 | 0.571 |
| SVM-RBF | 4 | 3 | 0 | 0.571 |
| XGB | 4 | 3 | 0 | 0.571 |
| GNB | 3 | 3 | 1 | 0.5 |
| LR | 0 | 0 | 7 | 0 |

We make the following observations:

1. For n=17 male subjects, Tree based models Random Forests (RF) and Decision Tree (DT) showed the highest diagnostic hit rate at 70% with 7 correct claims out of 10 total claims made against each of them.
2. For n=7 female subjects, Decision Tree (DT) had the highest diagnostic hit rate at 60% with 3 correct claims out of 5 total claims made against each of them.
3. Algorithms favoured by the authors of [10], notably Logistic Regression (LR), SVM and Naïve Gaussian Bayes (GNB) came second to the tree-based models for both genders. The authors explained that tree-based models are not suitable in such cases as they are likely to overfit. But we observe that for both genders, these algorithms had a lower ratio (around 1:1) of the number of correct diagnostic claims against incorrect ones compared to that for tree-based models using our LOO approach (almost ~2:1).

## V. CONCLUSION

This paper introduces a novel approach to leveraging shallow machine learning algorithms for voice pathology, specifically in how they can be used as a triaging tool to identify individuals at risk who would benefit from timely intervention. We also demonstrate the suitability of this method as an alternative to the constrained discrete non-continuous and intermittent data-capture methodologies using smartphone apps. The results presented provides reasonable validation as to how this innovative triaging system can be used alongside pre-existing commercial voice-based virtual assistants.

Many of these virtual assistants are now being adopted by healthcare institutions in telehealth and digital health supported pathways to improve patient experience, health outcomes and reduce pressure on the healthcare system. By embedding our triaging system in mainstream channels and solutions, we believe that the roll out, adoption and distribution of our system will be easier. Principally, we evaluated our concept and showed how it can be used as a continuous pre-diagnostic tool for older adult patients at home who may be undiagnosed or are pre-diabetic, without inconveniencing their daily life.

For our next work, we will evaluate the triaging system with a larger user cohort to study its generalisability and efficacy across co-existing medical conditions, age groups, and with alternative acoustic features like cepstral peak prominence. We foresee consecutive future works focusing on extending this voice-based triaging system for other medical conditions which can be managed better with early intervention like onset of cognitive decline, hypertension and respiratory tract infections.

# Authors' background

| Name | Prefix | Research Field | Email | Personal Email |
|------|--------|----------------|-------|----------------|
| Kelvin Summoogum | Associate Professor | Applied AI in Geriatrics | kelvin@miihealth.ai | kelvin@miicare.com |
| Debayan Das | Bachelor's Graduate | Acoustic Machine Learning, Clinical Data Science | debayan.das@miicare.co.uk | debayan.das.uk@gmail.com |
| Sathish Kumaran | Undergraduate Student | Clinical Data Analytics | sathish.kumaran@miicare.co.uk | sathishkumaran30@gmail.com |
| Dr. Sumit Bhagra, MD | N/A | Lead Physician & Chair of Endocrinology at Mayo Clinic | bhagra.sumit@mayo.edu | bhagra.sumit@mayo.edu |

**Note:**

[1] **This form helps us to understand your paper better; the form itself will not be published.**

[2] *Prefix*: can be chosen from Master Student, PhD Candidate, Assistant Professor, Lecture, Senior Lecture, Associate Professor, Full Professor